# Enhancing nonnative speech perception and production through an AI-powered application

**Georgios P. Georgiou[1 2]**

[1]Department of Languages and Literature, University of Nicosia, Nicosia, Cyprus

[2]Phonetic Lab, University of Nicosia, Nicosia, Cyprus

## Abstract

While research on using Artificial Intelligence (AI) through various applications to enhance foreign language pronunciation is expanding, it has primarily focused on aspects such as comprehensibility and intelligibility, largely neglecting the improvement of individual speech sounds in both perception and production. This study seeks to address this gap by examining the impact of training with an AI-powered mobile application on nonnative sound perception and production. Participants completed a pretest assessing their ability to discriminate the second language English /iː/–/ɪ/ contrast and produce these vowels in sentence contexts. The intervention involved training with the Speakometer mobile application, which incorporated recording tasks featuring the English vowels, along with pronunciation feedback and practice. The posttest mirrored the pretest to measure changes in performance. The results revealed significant improvements in both discrimination accuracy and production of the target contrast following the intervention. However, participants did not achieve native-like competence. These findings highlight the effectiveness of AI-powered applications in facilitating speech acquisition and support their potential use for personalized, interactive pronunciation training beyond the classroom.

## 1. Introduction

Segmental difficulties in both the perception and production of nonnative speakers are well-documented in the literature, often attributed to interference from their native language (Flege, 1993; Georgiou & Giannakou, 2024; Georgiou & Kaskamba, 2024; Navarra et al., 2005; Solon et al., 2017). However, nowadays, there is increasing recognition that age is not a limiting factor in acquiring native-like speech abilities in adulthood (Dąbrowska et al., 2020; Prela et al., 2024). The rapid advancement of Artificial Intelligence (AI) presents an opportunity to enhance learning outcomes in the often underexplored and marginalized area of pronunciation (Low, 2021). Pronunciation is an important subskill, playing a crucial role in second language communication, and is far from being insignificant (Levis, 2022). When integrated with other aspects of speech, such as vocabulary selection and grammar, it can have an even greater impact on comprehensibility (Ruivivar & Collins, 2018). Speech acquisition can move beyond traditional phonetic training toward more technologically advanced and engaging methods, enabling learners to refine their skills independently. This study examines the impact of an AI-driven pronunciation application on the speech perception and production skills of nonnative English speakers.

High Variability Phonetic Training (HVPT) is a widely adopted approach for enhancing listeners' and learners' speech perception and production skills (Almusharraf et al., 2024; Barriuso & Hayes-Harb, 2018; Georgiou, 2021a; 2022a; Mahdi et al., 2024; Saito et al., 2022). This method employs multiple speakers, rather than a single voice, and incorporates stimuli within various linguistic contexts. Georgiou (2022a) examined the impact of HVPT on the identification and production of English vowels in Cypriot Greek children and adults. The findings revealed significant improvements in both groups' perceptual abilities, with some transfer of benefits to production for children. However, the majority of studies have focused on HVPT within controlled laboratory settings. As Thomson (2011) proposed, HVPT is a promising candidate for integration into computer-assisted pronunciation training applications. The author highlighted that the development of a web-based platform would offer endless research possibilities, allowing teachers and researchers to collaborate remotely, monitor the effects of perceptual training on pronunciation, and refine the software. What Thomson envisioned nearly 15 years ago can now be realized through the use of AI technologies embedded in various types of applications.

The rapid expansion of AI in recent years has created significant opportunities for improving pronunciation learning (Georgiou, 2025). Learners can now engage with mobile and web applications, chatbots, and intelligent virtual assistants to enhance their pronunciation skills (Vančová, 2023). Research has highlighted the benefits of AI tools in various aspects of pronunciation, including comprehensibility, intelligibility, nativeness, and sound production (Chung & Bong, 2021; Chuyen et al., 2021; Sonsaat-Hegelheimer & Kurt, 2025; Yo & Ahn, 2024). For example, Sonsaat-Hegelheimer & Kurt (2025) examined the effect of practice with two different AI-based chatbots, Pronounce and Gemini, on L2 English learners' pronunciation comprehensibility. Pronounce provides explicit feedback, while Gemini offers implicit feedback through real-time transcription. Three groups of learners participated: two experimental groups using one of the chatbots and a control group. Although group-level comprehensibility ratings showed no significant improvement, individual learners demonstrated notable improvements.

Wang et al. (2025) investigated the effectiveness of SpanishBot, a chatbot designed to support nonnative speakers in mastering Spanish pronunciation. Participants were randomly assigned to either an experimental group using SpanishBot or a control group relying on independent learning methods. Over two months, the experimental group received personalized pronunciation feedback, leading to statistically significant improvements in pronunciation proficiency, including enhanced clarity, accuracy, and intonation of letter sounds and words. At the segmental level, Chuyen et al. (2021) investigated the effectiveness of the Duolingo application in improving Thai students' pronunciation of various English consonants. The results of the study indicated that six weeks after the intervention, the experimental group trained with Duolingo improved their pronunciation. In another recent study, Mingyan et al. (2025) explored the impact of an AI-powered mobile application on Chinese undergraduate EFL students' speaking performance. Using a quasi-experimental design, two groups were compared: one using WeChat for assignments (control group) and the other using both WeChat and Liulishuo (experimental group). The findings showed that the experimental group significantly outperformed the control group in pronunciation and fluency, though no significant improvements were found in vocabulary or grammar. The findings highlight the potential of AI-powered applications to enhance speaking skills in English as a foreign language.

An application designed for improving foreign language learners' pronunciation skills is Speakometer, which offers tools for practice and feedback. This mobile application utilizes advanced automatic speech recognition technology and AI to evaluate users' pronunciation skills and their ability to reproduce sounds and words accurately. With a database of over 2.000 English words and a system for practicing English sounds, users can track their progress by reviewing recorded audio and identifying pronunciation errors. Additionally, Speakometer provides access to an extensive database of over 65.000 words, each with pronunciation practice, helping users expand their vocabulary while improving pronunciation. Nguyễn (2024) explored the effectiveness of the Speakometer application in improving pronunciation among Vietnamese English teachers. Although there was limited pronunciation training and a focus on test-driven exams, participants using these applications showed significant improvement in second language sound perception compared to the control group, which used traditional methods. In contrast, Palomo Martín (2024) found that AI tools do not significantly improve students' pronunciation. The author examined the challenges faced by Spanish-Catalan learners and the potential of AI tools like Siri and Speakometer in pronunciation training. The study involved Spanish-Catalan teenagers, who were divided into two groups: one using traditional methods and the other using the Speakometer applications. The findings indicated that the AI tools did not substantially enhance pronunciation or result in a more native-like accent. One limitation of the above designs (i.e., an experimental group trained with AI and a control group trained with traditional methods) is the difficulty in controlling for confounding variables, such as participants' prior experiences or personal learning preferences, which could affect their progress in either the traditional or AI-based group.

Despite recent research on the effectiveness of AI applications in pronunciation learning, most studies have primarily focused on other aspects of pronunciation, rather than on the articulation of individual sounds. In addition, very few studies have examined the impact of AI-powered tools on the perception of speech sounds. The aim of this study is to investigate the effect of training using a mobile AI-powered application on the speakers' speech acquisition skills in a second language. The following research questions are to be answered: a) Do Cypriot Greek speakers of English as a second language improve their discrimination of the /iː/ and /ɪ/ contrast after training with the Speakometer application?; b) Do these speakers improve their production of the contrasts (in terms of F1, F2, and duration) after training? We chose Speakometer since it comprises a user-friendly mobile application, which explicitly targets the learning of English speech sounds (for a review of the application, see Kurt, 2022). The application uses an AI algorithm and automatic speech recognition to rate the users' pronunciation. Users can refine their pronunciation by incorporating AI-generated corrective feedback based on an analysis of their spoken input. We focus on the English /iː/–/ɪ/ contrast, which is problematic for Cypriot Greek speakers in both perception and production. This is due to the absence of the contrast members from the Cypriot Greek phonological inventory and the classification of both vowels in terms of the Cypriot Greek /i/

(Georgiou, 2019, 2022b). The findings of this study can offer significant insights into how AI-driven feedback can enhance second language speech acquisition and how such applications can effectively be integrated into foreign language practice.

## 2. Methodology

### 2.1 Participants

The study involved 20 Cypriot Greek female participants, aged between 20 and 29 ($M = 23.85$, $SD = 2.87$). All participants were graduate or postgraduate students at universities in Cyprus, with backgrounds from middle-income families. None had spent a significant amount of time in an English-speaking country. They reported beginning to learn English at the ages of 7–8. Participants had IELTS scores ranging from 6 to 7.5, which corresponds to the upper B2 and C1 levels. Additionally, all participants completed an online Cambridge English Test, confirming that they were within the B2–C1 range with scores between 15 and 22 ($M = 18.75$, $SD = 1.97$). For the control group in the perception task, eight female Standard Southern British English (SSBE) speakers were recruited, aged 22–37 ($M = 27.71$, $SD = 5.28$). The native productions for the production task, which involved 10 SSBE adult female speakers were obtained from Georgiou et al. (2024). All participants reported no language disorders and had normal vision and hearing.

### 2.2 Materials

The materials for the perception task included two monosyllabic real words, "heed" and "hid", each containing the vowels /iː/ and /ɪ/, respectively, in the /hVd/ context. Another eight words representing different vowels were used as filler words; these words were "head", "herd", "had", "hard", "hod", "hoard", "hood", and "who'd". All words were embedded within the carrier phrase "They say /hVd/ again" and were recorded by two adult female native speakers of SSBE in a quiet environment. A Zoom recorder at a 44.1 kHz sampling rate was used for the recordings; these were saved on a computer in 24-bit WAV format. The production task involved the same words within the carrier phrase presented in written form using standard British orthography.

### 2.3 Procedure

#### 2.3.1 Pretest

*Speech Perception Task*. Participants performed an AXB test to assess their ability to discriminate between the vowels /iː/ and /ɪ/. The test was scripted in Praat (Boersma & Weenink, 2025) and conducted in a quiet room. Specifically, participants listened from the computer loudspeakers to a triad of target words and were asked to select whether the middle vowel (X) was the same as the first (A) or the second vowel (B). All of them wore earphones throughout the procedure. The test included 32 items, consisting of 16 target contrasts, repeated four times across four configurations: AAB, ABB, BBA, and BAA, and 16 filler words. The X token was always a word produced by a different English speaker compared to A and B to avoid a purely auditory decision. The interstimulus interval was set to 1s, and the intertrial interval to 500ms. Before the main test, a four-trial familiarization phase was conducted to ensure participants understood the task. Each participant completed the task in approximately five minutes.

***Speech Production Task***. Participants independently read 16 sentences (2 vowels × 4 repetitions + 8 filler words) in a quiet room. The sentences were printed on paper and read aloud, recorded using a Zoom recorder. The order of the phrases was randomized for each participant. The target vowel words were isolated and analyzed in Praat, following standard acoustic analysis procedures, with settings for spectrogram analysis and vowel boundary detection. For these analyses, we applied the following settings: a positive window length of 0.025 s, a pre-emphasis of 50 Hz, and a spectrogram view range extending up to 5500 Hz. Frequency measurements began at the transition from the noise of the preceding /h/ into the onset of the vowel and ended at the transition from the vowel into the following /d/. To minimize the influence of adjacent sounds, frequencies were extracted at the vowels' midpoints. Vowel durations were manually annotated by the author, who identified the start and end points of each vowel token. The F1 and F2 values were normalized using the Bark scale.

## 2.3.2 Training

The participants received a one-month pronunciation training, consisting of four individual sessions, each lasting one hour. The first session was conducted in a laboratory setting. This enabled the researcher to familiarize participants with the application, assign them tasks and objectives, offer support for any challenges, and ensure they maintained consistent practice. The remaining three sessions were completed once per week at the participants' homes. During these sessions, participants used the Speakometer application on their smartphones to practice their pronunciation skills. They were trained in producing English monophthong vowels using the Speakometer session. They were presented with a list of 10 practice words for each vowel they selected. Each word appeared individually on the screen, accompanied by its phonetic transcription beneath the spelling. Participants could listen to a model pronunciation, and beneath the word and transcription, a microphone button allowed them to record their own pronunciation. After the recordings, a meter appeared on the screen, rating participants' accuracy on a four-point scale: red, orange, yellow, and green. Moreover, written feedback was displayed between the word and the meter. If the accuracy was poor, participants were prompted to re-record the word. At the end of the task, participants received an overall score reflecting their pronunciation abilities. The application automatically randomized words after each attempt across the different sessions. participants received a printed list of 20 words – 10 targeting the specific contrast and 10 addressing other contrasts – and they were instructed to locate these words using the application's dictionary and then pronounce them.. All participants were required to log their progress, take screenshots of their scores, and share them with the researcher. Figure 1 illustrates an example of pronunciation feedback given by the Speakometer application.

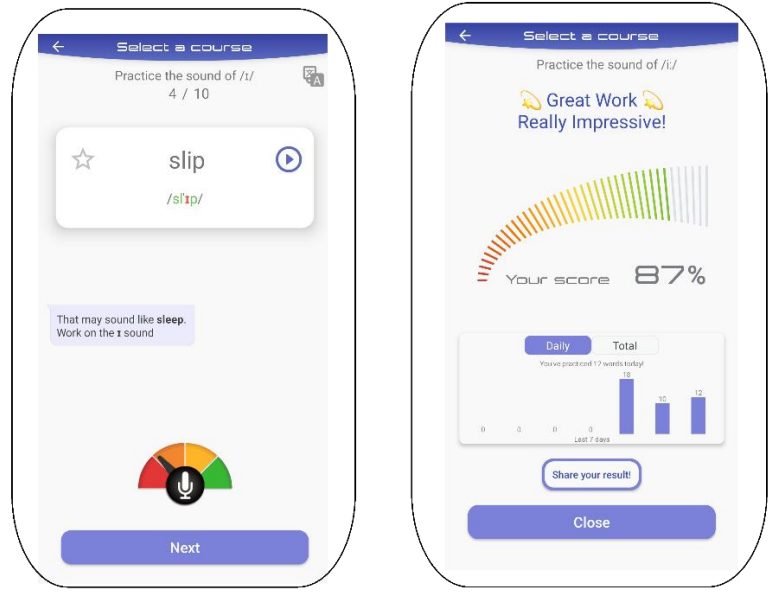

Figure 1: Example of pronunciation feedback given by the Speakometer application.

### 2.3.3 Posttest

One month following the intervention, participants completed the same speech perception and production tasks as in the pretest, allowing for comparison of results before and after the intervention.

### 2.4 Statistical Analysis

The results of the perception task were analyzed using a binomial mixed-effects model in R (R Core Team, 2025) with the lme4 package. Response (correct/wrong) was modeled as the dependent variable, Condition (prettest, posttest, native speakers) as the fixed factor, and Participant and Item as random factors. The results of the production task were analyzed by employing linear mixed-effects models from the lmerTest package. For the investigation of the distance of the two English vowels across all conditions, the vowels' Euclidean Distances (EDs) in the F1 × F2 vowel spaces were measured. ED is a widely used method in phonetic research for the measurement of the acoustic similarity or difference between vowels based on their formant frequencies. Specifically, it quantifies how far apart two vowels are in a two-dimensional space (Georgiou, 2021b; Georgiou & Dimitriou, 2023; Liu & Escudero, 2025). Condition was the fixed factor, and Participant was the random factor. To explore the duration of difference of the contrast, another linear mixed-effects model was utilized with Difference as the dependent variable. Similarly, Condition was set as the fixed factor and Participant as the random factor. Posthoc tests from the emmeans package were employed to examine differences across the conditions; the Tukey adjustment was used.

# 3. Results

## 3.1 Speech perception task

The percentages of correct responses across pretest, posttest, and native speakers are illustrated in Figure 2. Posttest results indicate improvement following the intervention, though perceptual abilities still differed from those of native speakers.

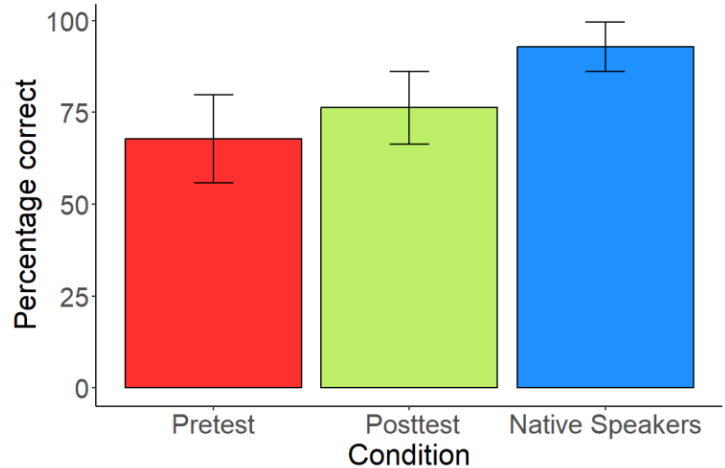

Figure 2: Percentage of correct responses in the perception task across the three conditions.

To examine potential differences across the three conditions, a generalized mixed-effects model was employed. The results indicated significant differences between ConditionPosttest, ConditionPretest, and the Intercept term. Further posthoc tests revealed significant differences between the responses of native speakers–posttest, native speakers–pretest, and pretest–posttest; these results show that speakers improved their discrimination accuracy after the intervention, although they did not acquire native-like competence. The findings of the statistical analysis are shown in Table 1.

Table 1: Results of the generalized mixed-effects model and the posthoc test

| Main model (intercept term: ConditionNativeSpeakers) | | | | |
|---|---|---|---|---|
| | Estimate | Std. Error | $z$-value | $p$-value |
| (Intercept) | 2.589 | 0.387 | 6.696 | < 0.001 |
| ConditionPosttest | –1.408 | 0.390 | –3.608 | < 0.001 |
| ConditionPretest | –1.834 | 0.387 | –4.744 | < 0.001 |
| Posthoc test (Tukey method) | | | | |
| Condition | Estimate | Std. Error | $z$-value | $p$-value |
| NativeSpeakers–Posttest | 1.408 | 0.390 | 3.608 | < 0.001 |
| NativeSpeakers–Pretest | 1.834 | 0.387 | 4.744 | < 0.001 |
| Posttest–Pretest | 0.426 | 0.179 | 2.385 | 0.045 |

## 3.2 Speech production task

As seen in Figure 3, the analysis of the speakers' vowel space demonstrated that English native speakers produced the English vowels /iː/ and /ɪ/ distinctly. In contrast, the nonnative speakers produced the two vowels in the pretest with high overlap. The overlap seems to be smaller in the posttest, with English vowel /ɪ/ of speakers being more back and moving towards the native counterpart.

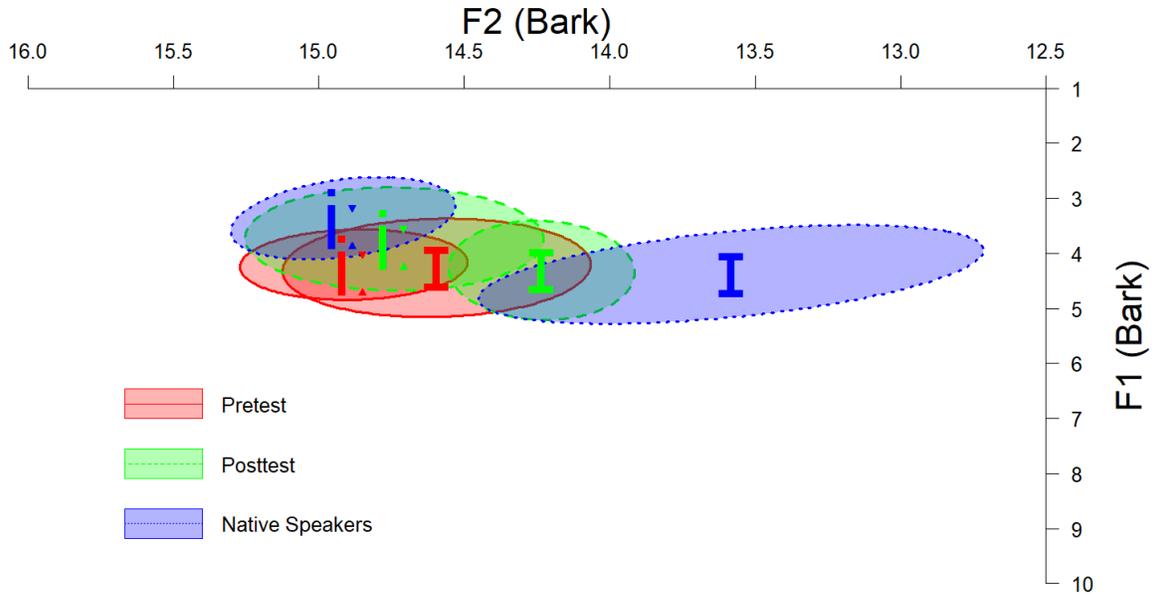

Figure 3: F1 × F2 of English vowels /iː/ and /ɪ/ across the three conditions.

A supervised machine learning analysis using R was employed to model the differences among the three conditions. The dataset contained a categorical variable, Condition. To ensure robust model evaluation, we partitioned the data into training (80%) and testing (20%) subsets using stratified sampling. We employed Linear Discriminant Analysis (LDA) (see also Georgiou 2023, 2024) to classify the two conditions using the predictor variable ED. The model was trained using five-fold cross-validation to optimize performance and mitigate overfitting. After training, the model was tested on the held-out test data. Predictions were obtained using the predict function, and classification performance was assessed through multiple metrics including accuracy, precision, recall, F1-score, and AUC/ROC curve (see Georgiou & Theodorou, 2024).

In terms of EDs, the model was highly effective at distinguishing native vs nonnative speech at the pretest stage compared to the posttest stage, suggesting that formant differences were less evident after the intervention. However, the low performance of the model at distinguishing speech at the pretest vs the posttest stages shows that some features still overlap between the two stages. In terms of duration differences, the model presented with low performance for posttest vs pretest, but with higher performance for the pretest/posttest vs native speaker condition. This suggests that no significant differences were observed in the duration contrast after the intervention. Although Figure 4 shows differences between the test conditions, the model's results may be influenced by the high variability that persisted in the posttest. The performance of LDA models is shown in Table 2.

Table 2: Performance of the LDA models

|  | Accuracy | Precision | Recall | F1-score | AUC/ROC curve |
|---|---|---|---|---|---|
| EDs |  |  |  |  |  |
| NativeSpeakers–Posttest | 0.77 | 0.67 | 0.83 | 0.74 | 0.85 |
| NativeSpeakers–Pretest | 0.80 | 0.81 | 0.93 | 0.87 | 0.91 |
| Posttest–Pretest | 0.65 | 0.69 | 0.65 | 0.67 | 0.62 |
| Duration difference |  |  |  |  |  |
| NativeSpeakers–Posttest | 0.80 | 1.00 | 0.80 | 0.89 | 0.55 |
| NativeSpeakers–Pretest | 0.75 | 0.94 | 0.79 | 0.86 | 0.78 |
| Posttest–Pretest | 0.59 | 0.69 | 0.57 | 0.63 | 0.68 |

A linear mixed-effects model was used for the statistical analysis. The test showed significant differences between ConditionPosttest, ConditionPretest, and the Intercept term. The posthoc test yielded significant differences across all conditions. So, EDs in the pre- and post-intervention conditions were different from the native condition, and ED in the pre-intervention condition was different from the post-intervention condition. Since the difference from the native condition is greater in the pre-condition than in the post-condition, this suggests some movement toward the native norm following the intervention. Results are presented in Table 3.

Table 3: Results of the linear mixed-effects model and the posthoc test for the EDs of the English contrast /iː/–/ɪ/

| Main model (intercept: ConditionNativeSpeakers) | | | | |
|---|---|---|---|---|
|  | Estimate | Std. Error | $z$-value | $p$-value |
| (Intercept) | 1.713 | 0.096 | 17.822 | < 0.001 |
| ConditionPosttest | –0.903 | 0.107 | –8.401 | < 0.001 |
| ConditionPretest | –1.044 | 0.107 | –9.717 | < 0.001 |
| Posthoc test (Tukey method) | | | | |
| Condition | Estimate | Std. Error | $z$-value | $p$-value |
| NativeSpeakers–Posttest | 0.903 | 0.108 | 8.401 | <0.001 |
| NativeSpeakers–Pretest | 1.044 | 0.108 | 9.717 | <0.001 |
| Posttest–Pretest | 0.141 | 0.057 | 2.492 | 0.039 |

The duration difference between English vowels /iː/ and /ɪ/ is smaller in the pretest compared to the duration difference in the posttest and native speaker conditions. The difference between the two vowels in the posttest and native speaker conditions is comparable (see Figure 4). A linear mixed-effect model indicated significant differences between ConditionPretest and the respective intercept term. The findings from the posthoc test demonstrated significant differences between native speakers–pretest and posttest–pretest. This means that after the intervention the duration difference for the contrast became native-like. The results of the analysis are presented in Table 4.

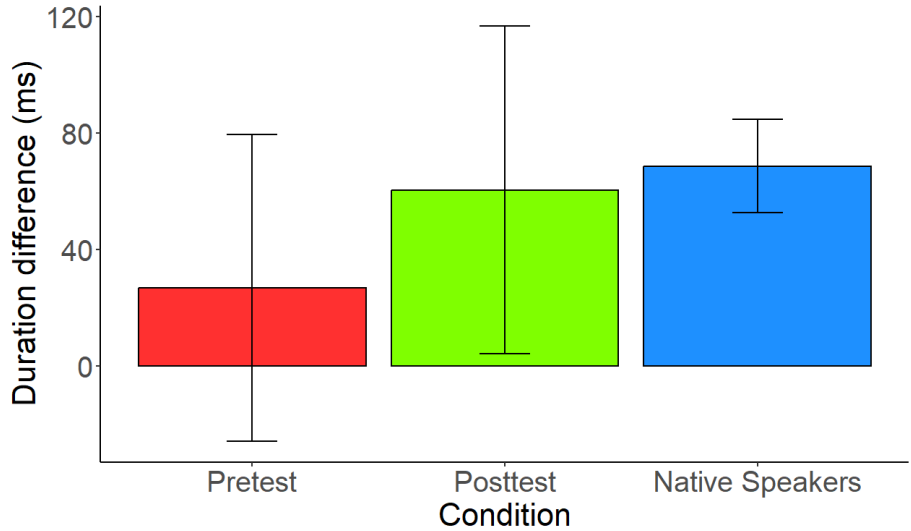

Figure 4: Duration difference between /iː/ and /ɪ/ across all conditions.

Table 4: Results of the linear mixed-effects model and the posthoc test for the difference in the duration of English /iː/ and /ɪ/

| Main model (intercept term: ConditionNativeSpeakers) | | | | |
|---|---|---|---|---|
| | Estimate | Std. Error | z-value | p-value |
| (Intercept) | 68.650 | 11.566 | 5.935 | < 0.001 |
| ConditionPosttest | −8.187 | 12.931 | −0.633 | 0.527 |
| ConditionPrettest | −41.839 | 12.931 | −3.235 | 0.001 |
| Posthoc test (Tukey method) | | | | |
| Condition | Estimate | Std. Error | t-value | p-value |
| NativeSpeakers–Posttest | 8.190 | 12.930 | 0.633 | 0.802 |
| NativeSpeakers–Pretest | 41.840 | 12.930 | 3.235 | 0.004 |
| Posttest–Prettest | 33.650 | 7.530 | 4.470 | < 0.001 |

## 4. Discussion

This study aimed to examine the impact of pronunciation training using an AI-powered web application on improving the perception and production of a challenging English vowel contrast. Participants underwent four one-hour training sessions with the mobile-based Speakometer application, which offers real-time pronunciation feedback and includes exercises for the improvement of speech perception. The pretest and posttest phases included tasks assessing both the discrimination of the English /iː/–/ɪ/ contrast and the controlled production of these vowels.

The first aim of this study was to investigate whether participants could improve their discrimination of the English /iː/–/ɪ/ contrast after receiving training through the mobile application. The results of the perception task showed that participants were able to perceive the contrast more accurately following the intervention. This aligns with the findings of other studies, which manifest the positive impact of applications like Speakometer on improving learners' speech perception skills (Nguyễn, 2024). Importantly, although the training was primarily production-based, it led to improvements in the speakers' perceptual abilities. This finding aligns with other studies indicating that production-based training can enhance the perception of speech sounds (Li, 2024; Stratton, 2025). The current findings support a bidirectional relationship between speech perception and production (Flege et al., 2021; Georgiou & Savva, 2025) and also contribute to the ongoing debate on which domain develops first, suggesting that production may precede perception (Raver-Lampman & Wilson, 2018). Another observation is that speakers did not achieve native-like perceptual accuracy (see also Georgiou, 2021a). This outcome was expected, as the participants only completed four sessions of training using the application. It is possible that a longer-term training regimen could have resulted in even greater improvements in the participant's performance on the speech perception task.

The second objective of this study was to examine the effect of training on the production of the target contrast, focusing on the first two formant frequencies and the duration of the English /iː/–/ɪ/ vowels. The results revealed a significant difference in the EDs of the contrast between the pretest and posttest phases, with the post-intervention distance shifting closer to the native-speaker norm. This suggests that training influenced the participants' articulatory patterns, leading to improved production performance. Moreover, the duration difference between /iː/ and /ɪ/ increased after training, becoming more native-like. This may indicate that learners internalized temporal cues as an essential phonetic distinction in English pronunciation. However, it is important to acknowledge that the duration differences in the posttest exhibited high variability among participants, suggesting that improvement was not uniform across all speakers. Future research should explore whether extended training or more personalized feedback could help reduce variability and lead to more consistent gains in pronunciation. Overall, these findings reinforce previous research highlighting the effectiveness of AI-driven applications in enhancing pronunciation accuracy (Chuyen et al., 2021; Khampusaen et al., 2023; Wang et al., 2025). However, as observed in the perception task, participants' production skills did not reach native-like levels, which was expected given the relatively short duration of training.

The findings provide some practical implications for pronunciation training. The success of AI-powered training in improving speech acquisition performance highlights the potential of such applications for second language learners, particularly in self-directed learning environments (Qiao & Zhao, 2023). Laboratory-based HVPT is another effective approach, exposing learners to a wide range of speech variations by using multiple speakers and contexts during training. This method has been shown to improve learners' ability to generalize phonetic distinctions across different speakers and situations (Barriuso & Hayes-Harb, 2018). While HVPT is effective in improving speech acquisition, AI-powered pronunciation training offers a more personalized, interactive, and flexible approach. AI tools provide real-time feedback, adaptive learning paths, and self-directed practice, making them especially useful for learners who need continuous, accessible pronunciation training outside of structured classroom environments. AI tools could be integrated into language courses as supplementary pronunciation training resources, helping learners fine-tune their speech in ways that traditional classroom instruction might not fully support (Young, 2024). Finally, the observed variability in the production task implies that some learners may need additional support, such as explicit phonetic instruction alongside AI-based practice, to maximize their pronunciation gains.

# 5. Conclusions

This study demonstrates the potential of AI-powered mobile applications to facilitate the acquisition of difficult second language contrasts, particularly by improving both perception and production. Given that pronunciation training is often overlooked in second language instruction due to time constraints, lack of teacher expertise, or insufficient resources, the findings show that AI-based tools can provide an accessible and effective alternative for learners. Moreover, this study stands out as one of the few that directly examines the effect of AI-driven pronunciation training on both perceptual and productive aspects of specific sound contrasts. By demonstrating measurable improvements in nonnative speakers' abilities, this study contributes to the ongoing shift toward technology-enhanced pronunciation training and opens the door for further exploration of AI's role in second language acquisition.

While the study presents promising results, it also has some limitations that should be addressed in future research. It specifically examined the /iː/–/ɪ/ contrast, which is known to be challenging for Cypriot Greek learners of English. However, different learners struggle with different phonetic contrasts depending on their first language. Future research should investigate whether AI-powered pronunciation training can be equally effective for a broader range of phonemic contrasts. Furthermore, participants only underwent four training sessions, which, while sufficient to yield noticeable improvements, may not have been long enough to establish long-term learning effects. Longer-term studies are needed to determine whether extended training leads to more robust and lasting improvements in production and perception. While immediate posttests showed gains in speech acquisition, it remains unclear whether these improvements persist over time. Future studies should incorporate a delayed posttest (e.g., after several weeks or months) to assess the retention of learning. This would help determine whether AI-driven pronunciation training leads to permanent changes in learners' phonological representations or if the improvements fade without continued practice. Finally, to better understand the effectiveness of AI-powered pronunciation training, future research should compare it to other training methods, such as explicit perception training, traditional classroom instruction, or hybrid approaches that combine AI with teacher-led feedback.

## Competing interests

## Ethical approval

## Informed consent

## Acknowledgments

This study is supported by the Phonetic Lab at the University of Nicosia. We would like to thank all the participants for their involvement.

## Data availability

The datasets generated during and/or analyzed during the current study are available from the corresponding author on reasonable request.

Ruivivar, J., & Collins, L. (2018). Nonnative accent and the perceived grammaticality of spoken grammar forms. *Journal of Second Language Pronunciation, 5*, 270–294.

Saito, K., Hanzawa, K., Petrova, K., Kachlicka, M., Suzukida, Y., & Tierney, A. (2022). Incidental and multimodal high variability phonetic training: Potential, limits, and future directions. *Language Learning*, *72*(4), 1049–1091.

Solon, M., Long, A. Y., & Gurzynski-Weiss, L. (2017). Task complexity, language-related episodes, and production of L2 Spanish vowels. *Studies in Second Language Acquisition*, *39*(2), 347–380.

Sonsaat-Hegelheimer, S., & Kurt, Ş. (2025). The impact of generative AI-powered chatbots on L2 comprehensibility. *Journal of Second Language Pronunciation*.

Stratton, J. M. (2025). The effects of production training on speech perception in L2 learners of German. *Journal of Phonetics*, *108*, 101370.

Thomson, R. I. (2011). Computer assisted pronunciation training: Targeting second language vowel perception improves pronunciation. CALICO *Journal, 28*(3), 744–765.

Vančová, H. (2023). AI and AI-powered tools for pronunciation training. *Journal of Language and Cultural Education*, *11*(3), 12–24.

Wang, Y. F., Hsu, M. H., Wang, M. Y. F., & Chang, Y. T. (2025). Enhancing phonetic accuracy through chatbot-assisted language learning. *Educational technology research and development*, 1–17.

Yoo, S., & Ahn, H. (2024). The Effects of Prosody Training with AI Chatbot on the English Pronunciation Improvement of Korean EFL Learners. 영어학, *24*, 1300–1317.

Young, C. J. (2024). *Evaluation of speech recognition, text-to-speech, and generative text artificial intelligence for English as a foreign language learning speaking practices* (Doctoral dissertation). Tokyo Denki University.
16